# Machine Learning-based Approach for Ex-post Assessment of Community Risk and Resilience Based on Coupled Human-infrastructure Systems Performance

Xiangpeng Li[1], Ali Mostafavi[2]
[1] Ph.D. student, UrbanResilience.AI Lab, Zachry Department of Civil and Environmental Engineering, Texas A&M University, College Station, TX, 77843; e-mail: xplli@tamu.edu
[2] Associate Professor, Urban Resilience.AI Lab Zachry Department of Civil and Environmental Engineering, Texas A&M University, College Station, TX, 77843; e-mail: amostafavi@civil.tamu.edu

## Abstract

While current approaches primarily focus on anticipatory or predictive resilience assessments of natural events, there is a limitation in the literature of data-driven analyses for the ex-post evaluation of community risk and resilience, particularly using features related to the performance of coupled human-infrastructure systems. To address this gap, in this study we created a machine learning-based method for the ex-post assessment of community risk and resilience and their interplay based on features related to the coupled human-infrastructure systems performance. Utilizing feature groups related to population protective actions, infrastructure/building performance features, and recovery features, we examined the risk and resilience performance of communities in the context of the 2017 Hurricane Harvey in Harris County, Texas. These features related to the coupled human-infrastructure systems performance were first processed using the PCA to get the risk and resilience index. Then, Hierarchical Clustering method was applied to classify census block groups into four distinct clusters then, based on the values of two indexed, these clusters were labeled and designated into four quadrants of risk-resilience archetypes. Finally, we analyzed the disparities in risk-resilience status of spatial areas across different clusters as well as different income groups. The findings unveil the risk-resilience status of spatial areas shaped by their coupled human-infrastructure systems performance and their interactions. The results also inform the feature disparities among four archetypes. For example, in high-risk areas, total building damage amount contributed a lot, and non-essential activity recovery contributed a lot in greater resilience. Our results also underscore the complex relationship between socio-economic factors, risk exposure, and resilience. Notably, clusters with higher income levels exhibit a positive association between income and resilience in high-risk settings, suggesting that these communities may benefit from more substantial investments in resilience resources. In contrast, lower-income clusters show a negative interaction effect between risk and income on resilience, implying that income alone is insufficient to mitigate the resilience challenges posed by elevated risk. This differential impact underscores the importance of considering both socio-economic and risk factors to develop targeted resilience strategies across varying community archetypes. The outcomes of this study provide researchers and practitioners with new data-driven and machine intelligence-based methods and insights to better evaluate the risk and resilience status of communities during a disaster to inform future and policies.

## 1. Introduction

Most of the existing work into community resilience to disasters and crises focuses on ex-ante (before-the-event) assessment of risk and resilience, with limited attention devoted to ex-post (after-the-event) assessment based on the actual performance of communities' coupled human-infrastructure systems. Ex-post community risk and resilience assessment involves examining and drawing conclusions about the extent of hazard impact and how effectively the coupled human-infrastructure systems performed in coping with the impacts, providing key insights for future resilience planning and strategy development (Molinari et al., 2014; Kameshwar et al., 2019). The current literature primarily emphasizes anticipatory, predictive, or ex-ante resilience assessments conducted before hazard events (Beguería, 2006; Wang et al., 2015; Yuan et al., 2022). Such analyses pay little attention to the resilience analysis at the post-event level, failing to capture, specify, and evaluate the actual status of community risk and resilience in an actual event context. In addition, the current approach to post-disaster assessments, as noted by Brown et al. (2008) and Mills et al. (2011), often relies heavily on survey methods, which are subject to information delays and data collection lags and suffer from major limitations in capturing the ex-post community performance. Another major limitation in the current approaches to ex-post risk and resilience assessments is their focus on their limited number of coupled human-infrastructure systems features. For example, some studies (Vamvakeridou-Lyroudia et al., 2020; Chen et al., 2016) focus primarily on hazard impacts, and some focused on digital twin with graphs (Hofmeister et al., 2024), while other studies (Pant et al., 2018; Kasmalkar et al., 2020) primarily focus on infrastructure disruptions or risk analysis (Najafi, Zhang, & Martyn, 2021, Han et al., 2024, Zheng et al., 2022). There is limited literature that examines the interplay between risk and resilience features in the context of ex-post big data analysis. Specifically, few studies have analyzed how individual risk and resilience features contribute to the overall risk and resilience levels of communities (Bertilsson et al., 2019). This gap in research limits the ability to provide comprehensive insights into the factors driving vulnerability and recovery.

The emergence of large-scale data affords an unprecedented opportunity to use a broad range of features related to the coupled human-infrastructure systems performance in ex-post assessment of risk and resilience across different areas of communities. This data-driven method offers immediate understanding of risk-resilience archetypes, enhancing our ability to perform ex-post evaluation of community risk and resilience. Flood risk has been extensively studied, and it can generally be categorized into two key aspects (Plate, 2002). The first focuses on managing existing systems, which include various processes aimed at mitigating flood disasters. This involves actions such as improving maintenance, enhancing preparedness, and organizing evacuation plans to ensure a rational and proactive approach to flood risk management. The second aspect pertains to disaster response, which includes activities like rescue operations during or immediately after a flood event. Risk perceptions are frequently regarded as a crucial element in vulnerability assessments (Birkholz, Muro, Jeffrey, & Smith, 2014, Ekmekcioğlu, Koc, & Özger, 2022). Resilience can broadly be defined as the capacity of an individual, community, city, or nation to withstand, absorb, or recover from an event—such as an extreme flood—or to effectively adapt to adverse conditions or changes, such as climate change or economic challenges, in a timely and efficient way (Sayers et al., 2013). Therefore, in this study, we will divide the features into risk and resilience features to analyze.

This study focuses on conducting a data-driven machine learning-based ex-post assessment of community risk and resilience based on the interplay among coupled human-infrastructure systems performance features. To this end, using data from the 2017 Hurricane Harvey in Harris County, Texas, this study captures three components of coupled human-infrastructure systems performance (Figure 1). Using a K-means clustering method, we classified spatial areas (census block groups (CBGs)) based on their feature similarity. The obtained clusters are then analyzed based on their features patterns to specify archetype labels for each cluster based on their risk and resilience status. The clusters are categorized into four archetypes: high-risk/high-resilience (HH), low-risk/high-resilience (LH), high-risk/low-resilience (HL), and low-risk/low-resilience (LL), based on an in-depth analysis of the coupled human-infrastructure systems features in each cluster (Figure 1). This approach unveils the risk and resilience of areas of the community based on the coupled human-infrastructure systems performance related to protective actions, infrastructure/buildings disruptions, and population recovery. Specifically, the results reveal that different spatial areas exhibit varying levels of risk and resilience, confirming the existence of four archetypes of risk and resilience status.

Enhancement of community resilience processes hinges on both ex-ante (anticipatory) and ex-post (observational) assessments. This study addresses an important gap related to the dearth of data-driven approaches for ex-post community risk and resilience assessment. The contributions of this study are fourfold. First, unlike previous approaches that assessed ex-post resilience based on infrastructure performance. This study captures heterogeneous features related to the observed performance of the components of coupled human-infrastructure systems. Second, the computed features related to the performance of coupled human-infrastructure systems (i.e., protective actions, infrastructure/building impacts, and population activity recovery) are based on novel data sources, enabling a data-driven approach for ex-post rather than survey-based methods. Third, the use of a machine learning approach enables classifying risk and resilience of spatial areas based on the similarity of their coupled human-infrastructure systems performance features. Accordingly, the machine learning-based approach addresses the limitations of index-based methods that rely on subjective feature weights. Fourth, evaluation of the patterns gives insight into features that shape risk and resilience patterns, providing knowledge for informing future plans and actions. These contributions address the current limitations in methods for ex-post risk and resilience assessment and offer new methods and insights to interdisciplinary researchers and practitioners across disaster science, urban science, and emergency management for ex-post assessment of community risk and resilience through data-driven and machine intelligence-based methods.

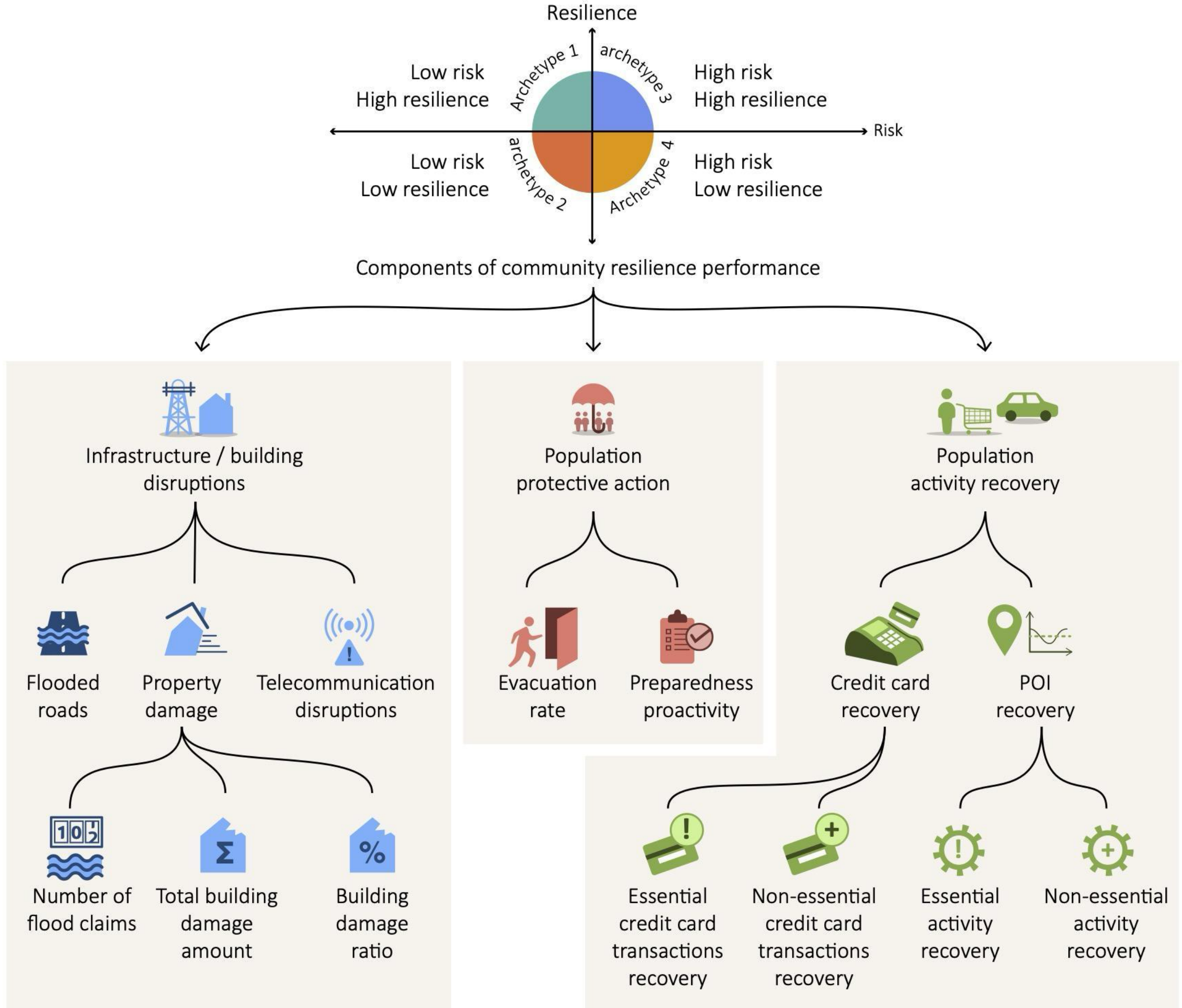

**Figure 1. Characterization of community resilience based on coupled human-infrastructure systems performance.** Coupled human-infrastructure systems performance are analyzed based various features: infrastructure/building disruptions population protective action, and population activity recovery. Each feature category has sub-components for total of 11 features. Based on the coupled human-infrastructure systems performance, spatial areas are grouped into four clusters of risk and resilience archetypes: high-risk/high-resilience (HH), low-risk/high-resilience (LH), high-risk/low-resilience (HL), and low-risk/low-resilience (LL)

## 2. Coupled Human-infrastructure Systems Performance Features

In this study, based on the existing literature, we focus on three components of the coupled human-infrastructure systems performance that shape community risk and resilience (Figure 1): population protective actions, infrastructure/building disruptions, and population activity recovery. Protective actions, particularly preparedness and evacuation, play a crucial role in community resilience (Dong et al., 2021). Preparedness and evacuation actions moderate the extent of harm people experience and how they cope with impacts in a hazard event (Dong et al., 2019). The effectiveness of people's preparedness actions is directly linked to their ability to stay shelter-in-place and cope with the impacts of hazards (Bullock et at., 2017; Lindell et al., 2006; Bronfman et al., 2019). As part of preparedness action, people would visit critical facilities to purchase necessary items for hazard preparation, including visits to grocery stores, gas stations, and other points of interest (POIs) (Dargin et al., 2021). Accordingly, we included a preparedness feature as one of the protective action features in our study. Preparation for evacuation, with its primary objective being to minimize the risk of loss of life or injury in the case of a disaster event, is primary protective action, serves as a crucial life-saving measure, (Cova et al., 2009; Lee et al., 2022).

The second component of the coupled human-infrastructure system performance is infrastructure/building disruptions. In this study, we captured features related to road inundations, disruptions in telecommunication services, and building damage to capture infrastructure/building disruptions. The rationale for focusing on these features were the extensive disruptions to roads and buildings caused by Hurricane Harvey flooding, as well as disruptions in telecommunications services. Other infrastructure disruptions (such as power outage) were not considered since Harvey did not cause extensive power outage in Harris County. Notably, Hurricane Harvey had a profound impact on transportation, such as road inundation and accessibility, with its devastating effects lingering for weeks after the storm (Rajput et al., 2022; Coleman et al., 2020). In addition, building damage was a key factor in household displacement and dislocation during Harvey (Rosenheim et al., 2021). Hurricane Harvey also caused disruptions in telecommunication services. Spatial analyses indicate that hazards can cause internet disruptions even in unaffected areas, highlighting the complex relationship between hazard severity and internet service continuity (Gupta et al., 2023).

The third component of the coupled human-infrastructure system performance was population activity recovery. The speed at which affected populations resume their normal life activities has been shown in prior studies to provide an important indicator for community recovery (Jiang et al., 2022). Population activities are considered to recover when people settle into a pre-disaster lifestyle after coping with impacts, disrupted infrastructure is restored, and businesses resume operations. For capturing population activity recovery features, we compared variations in visits to points of interest after the hazard event and with those normal period trends against which we measured duration and speed of recovery. Prior research (Podesta et al., 2021) has shown the effectiveness of evaluating fluctuations in visits to POIs for measuring and quantifying population activity recovery.

The components of the coupled human-infrastructure systems performance discussed above were captured using features computed from various datasets. Community-scale big datasets allow us to observe the dynamics of coupled human-infrastructure systems in the risk and resilience index of communities after hazard events. Harnessing community-scale big data is instrumental in the process of enhancing predictive flood risk monitoring, quick impact assessment, and situational awareness (Yuan et al., 2022, Praharaj et al., 2021). Multiple aspects of coupled human-infrastructure systems performance can be captured from community-scale big data and used to evaluate community risk and resilience status after the fact. In the following sections, we describe the datasets and methods used to determine the above-mentioned features.

## 3. Study Context and Datasets

### 3.1 Study Context

Beginning as a tropical storm in early August 2017, Hurricane Harvey rapidly escalated to a Category 4 hurricane. By August 23, it had intensified, necessitating mandatory evacuations in several Texas counties as forecasts predict the storm as to escalate into a major hurricane (Mirbabaie et al., 2020). Making landfall about 50 kilometers (about 30 miles) northeast of Corpus Christi, Texas, on August 26. Harvey then moved towards Houston, causing severe and widespread flooding in southeastern Texas, with particularly devastating impacts occurring on August 27. These conditions persisted until August 30, 2017, and Harvey was officially declared over on August 31, leading to widespread recovery and reconstruction efforts across numerous Texas and Louisiana counties (Sternitzky DiNapoli, 2017).

From August 25 through 30, 2017, Hurricane Harvey drenched Harris County with more than 130 centimeters (50 inches) of rain, causing unprecedented flooding across vast areas of the city (Emanuel, 2017). Even though the hurricane's center bypassed Houston to the south, it brought significant rains and floods to the area, particularly in the northeast, due to a stationary front around the storm (Scientific Investigations Report, 2018). Harris County, encompassing the densely populated Houston area, experienced more than 100 centimeters (40 inches) of rain, resulting in considerable flooding and damage (Qin et al., 2020; US Census Bureau, 2022). Harvey's ferocious winds, peaking at 150 miles per hour, caused localized damage and necessitated emergency actions to avert dam failures, further aggravating the flooding and infrastructure destruction (Blake & Zelinsky, 2018; Kiaghadi & Rifai, 2019). The significance of impacts, as well as the breadth of geographic area that was impacted makes this study context a suitable setting for ex-post assessment of community risk and resilience in this study.

### 3.2 Datasets for Features Calculation

The data sources utilized in this study are summarized in Table 1. Our analysis focused on computing the coupled human-infrastructure systems features at the CBG level. As shown in Table 1, the components of the coupled human-infrastructure systems performance we examined include three main feature groups: population protective action, infrastructure/building disruptions and population activity recovery. Population protective action, encompassing preparedness proactivity and evacuation rate, are calculated based on location-based data. Infrastructure/building disruptions include flooded roads, property damage, and telecommunications disruptions. Population activity recovery durations were computed based on location-based data (POI data) and credit card transactions. In total, the analysis examined 11 features related to the coupled human-infrastructure systems performance: flooded roads, number of flood claims, total building damage amount, building damage ratio, telecommunication disruptions, preparedness proactivity, evacuation rate, essential activity recovery, non-essential activity recovery, essential credit card transaction recovery, and non-essential credit card transaction recovery. There one socio-demographic characteristic feature is used to analyze the research result. A detailed description of each feature and the datasets and methods used in computing the features are presented below.

**Table 1. Coupled Human-infrastructure Features and Datasets**

| Feature Group | Feature | Data Description | Data Sources |
| --- | --- | --- | --- |
| Socio-demographic characteristic | Median income | Household median income in 2017 | U.S. Census Bureau table data |
| Protective action features | Preparedness proactivity | Maximum POI visits during preparedness period relative to normal baseline in pharmacy | Safegraph, Inc; Spectus; Microsoft building footprint |
| | | Maximum POI visits during preparedness period relative to normal baseline in gas station | Safegraph, Inc; Spectus; Microsoft building footprint |
| | Evacuation rate | Maximum evacuation rate from August 1 to September 31 relative to normal baseline at CBG level | Spectus |
| Infrastructure and building disruptions | Flooded roads | Total distance that has null traffic speed | INRIX |
| | Number of flood claims | The total count of flood claims within each census block group area | Federal Emergency Management Agency National Flood Insurance Program |

| | Total building damage amount | The actual cash value of the damage that the main property sustained is shown in terms of whole dollars. | Federal Emergency Management Agency National Flood Insurance Program |
|---|---|---|---|
| | Building Damage Ratio | The building damage ratio was determined by dividing the total number of claims by the total number of buildings in a given area. | Federal Emergency Management Agency National Flood Insurance Program, and Microsoft Building Footprints |
| | Telecommunication data | Maximum download kbps changing rate during hazard relative to normal baseline | Ookla |
| Recovery features | Essential activity recovery | Essential human mobility recovery data (essential POI) | Spectus |
| | Non-essential Activity Recovery | Non-essential human mobility recovery data (non-essential POI) | Spectus |
| | Essential credit card transaction recovery | Essential credit card activity | Safegraph |
| | Non-essential credit card transaction recovery | Non-essential credit card transactions recovery | Safegraph |

- *Preparedness proactivity*

We employed the method of preparedness proactivity from Li and Mostafavi (2022) to capture the earliest maximum POI change percentage (Equation 1) at CBG level using location-based data from Spectus. Spectus acquires its data through partnerships with smartphone applications, collecting information from devices where users have consented to the collection of their location data. Mobile phone data offer detailed insights into human movement patterns on a large scale at an unprecedented spatio-temporal granularity and scale (Yabe et al., 2022). By collaborating with app developers, Spectus harnesses a variety of signals, including Bluetooth, GPS, Wi-Fi, and IoT, to compile a high-resolution geo-behavioral dataset. In our study, the preparedness period was from August 20 through August 25, 2017. To assess preparedness proactivity, we identified the peak visitation percentage change date for each CBG relative to the baseline and then determined the interval between this peak and the hurricane's landfall. Greater preparedness proactivity means people in the community prepare earlier for the natural hazard. Zero proactivity indicates people reach maximum POI change percentage on the day of the hurricane landfall. To determine the changes in POI visits due to hurricane preparations, we established baselines using POI visitation patterns related to the first two weeks in August. This period's visitation numbers capture the residents' normal POI visits with no disturbances. The preparedness rate is calculated based on visitations to pharmacies and gas stations, determined using datasets of POIs visits. Equation 1 describes the percent change of the POI visits in calculating the preparedness index of each CBG:

$$PC_{i,d,t} = \frac{V_{i,d,t} - B_{i,d,t}}{B_{i,d,t}} \qquad \text{(Eq.1)}$$

where, $PC_{i,d,t}$ is the percentage change of visits to one category of POI (t) from home CBG (i) in date (d), $V_{i,d,t}$ is the number of visits to one category of POI (t) from home CBG (i) on date (d). $B_{i,d,t}$ is the calculated baseline value corresponding to the date. The maximum POI visitation change percentage of certain POI is computed, and the date is recorded. The preparedness proactivity is calculated based on the difference in days between the recorded date and August 25, 2017. For example, if an area reached its maximum POI visitation change percentage on August 24, 2017, its preparedness proactivity is one day. After compiling all preparedness proactivity measures related to gas stations and pharmacies, we determine their mean value as the overall preparedness proactivity for a CBG.

- *Evacuation Rate*

For each CBG, we calculated the maximum evacuation rate by comparing evacuation rate just before, during, and after Hurricane Harvey (August 6 through September 30, 2017) with a pre-event baseline period (July 9, through August 5, 2017). We then chose the maximum percentage change for each CBG as our feature value, following the methodology of Lee et al. (2022), and using location-based data from Spectus. Only users with a minimum of 240 minutes of daily location information were included to ensure data accuracy and minimize bias. Individuals who left their home CBGs and CBGs and stayed in another CBG for at least 24 hours during Hurricane Harvey were considered to have evacuated. Here Equation 2 describes the percent change of the evacuation rate (ER):

$$ER\ Change\ rate = \frac{ER_t - ER_B}{ER_B} \qquad \text{(Eq. 2)}$$

$ER_t$ represents the evacuation rate on day t, and $ER_B$ is the baseline. The maximum evacuation rate change percentage is the feature for evacuation in this study.

- *Flooded Roads*

Flooded road networks limit access to emergency services (Yuan et al., 2022; Dong et al., 2020; Bhavathrathan & Patil, 2015). Harris County Road segment traffic data from August 20 through September 11, 2017, obtained from INRIX, provided average traffic speeds for each road segment at 5-minute intervals, along with historical average speeds for comparison. Following the approach of Fan et al. (2020) and Yuan et al. (2021b, 2021c), road segments displaying a null value for average traffic speed were categorized as flooded during Hurricane Harvey. Using mapping tools, we created Line String representations of flooded roads to visualize their paths. By overlaying these lines with CBG boundaries, we identified where flooded roads were encompassed by a CBG polygon. We quantified the total length of flooded roads within each CBG polygon. These lengths represent the total null distance in each CBG polygon, indicating the extent of flooded roads within those areas.

- *Number of flood claims*

The insurance claim data utilized in this study was sourced from the FEMA National Flood Insurance Program (NFIP) (FEMA, 2023). FEMA manages the NFIP by overseeing the proper processing of insurance applications, determining accurate insurance premiums, and handling the renewal, modification, and cancellation of insurance policies. This data set provides details on NFIP claims transactions, and the number of claims in a given area was determined by calculating the total count of insurance claims filed by policyholders in each CBG.

- *Total building damage amount*

The total building damage amount, sourced directly from the National Flood Insurance Program, represents the actual cash value of damage incurred by the main property, expressed in whole dollars. We aggregated this data to the total building damage amount for each CBG area by summing all the reported damage amounts.

- *Building damage ratio*

The building damage ratio was determined by dividing the total number of claims by the total number of buildings in a given area. The number of claims has been mentioned before, and the data on the total number of buildings was from Microsoft Building Footprints. The Bing Maps and Microsoft Maps & Geospatial teams have extracted a comprehensive dataset consisting of 129,591,852 building footprints across the United States (*Microsoft/USBuildingFootprints*, 2018/2024).

- *Telecommunication disruptions*

Ookla data encompasses to cellular internet speeds, including specific metrics like upload and download rates from mobile devices, latency, and location-based information about both the device and the server (Ookla 2023). In our research, we focus on using the download kbps changing rate as the telecommunication data from August 10 to September 8, 2017. The download kbps changing rate was assumed to be the maximum changing rate during this period. For our analysis, we leveraged data provided by Ookla, which offers comprehensive insights into internet capabilities, including cellular internet speeds. This dataset includes vital metrics such as upload and download rates from mobile devices, latency, and geolocation information about the devices and servers involved (Gupta et al., 2023).

In our specific study, in order to record the maximum telecommunication disruptions, we concentrate on the maximum changing rate of download speeds in kilobits per second (kbps) from August 10 through September 8, 2017. To establish a baseline for normal telecommunication operations, we calculated the average download speed during the first two weeks of August. This period is chosen to reflect a typical pattern of telecommunication use without the influence of an impending natural disaster. The measure of the maximum telecommunication disruption is then determined by the Equation (3).

$$Maximum\ Download\ Spped\ Change\ Rate = \frac{Speed_B - min\ (Speed_t)}{Speed_B} \quad \text{(Eq.3)}$$

where $min\ (Speed_t)$ represents the minimum download speed on day t from August 24th and September 3rd, 2017, and $Speed_B$ is the baseline.

- *Essential and non-essential activity recovery and credit card transaction recovery*

To create smart, resilient, and sustainable urban environments, urban planned need an understanding of how extreme weather events affect human activities (Chen et al., 2020; Zhang et al., 2022). creating. The dataset comprising points of interest visitation patterns was obtained from Spectus; the dataset related to credit card transactions was sourced from SafeGraph. In line with methodologies used in prior research (Podesta et al., 2021; Li et al., 2023; Jiang et al., 2022), this study uses essential activity recovery, non-essential activity recovery, essential credit card transaction recovery, and non-essential credit card transaction recovery (Figure 2).

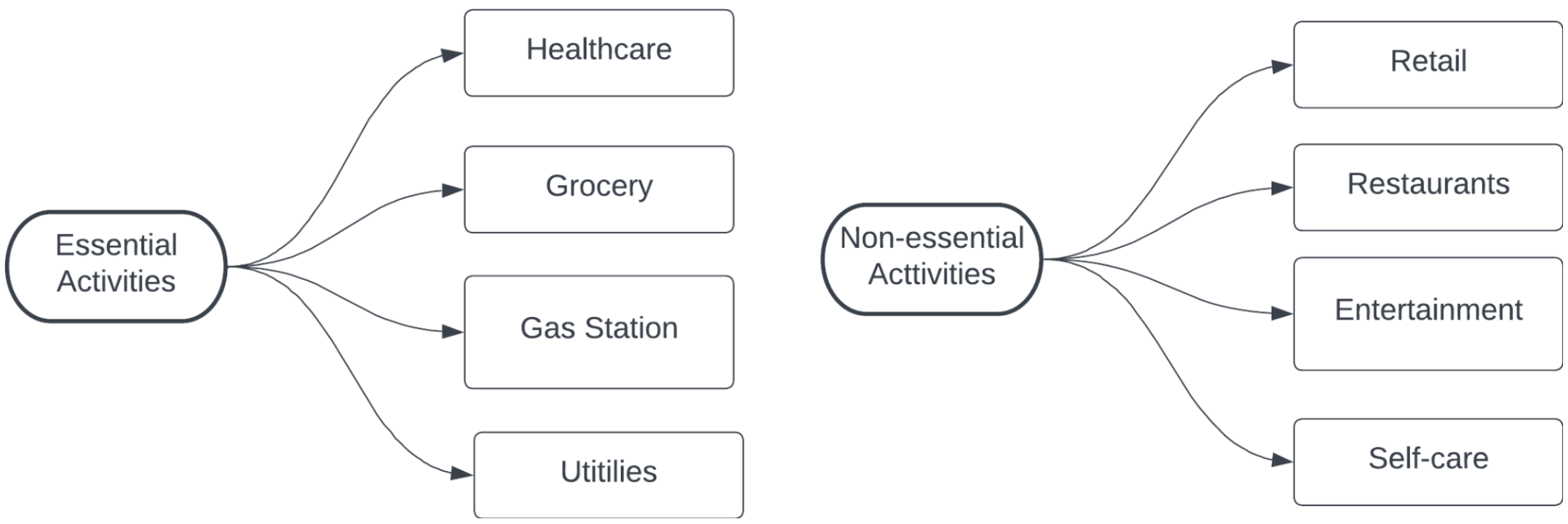

**Figure 2. Essential activities, and non-essential activities.** Essential activities represent services that are critical for daily life and survival, such as healthcare, grocery shopping, gas stations, and utilities. In contrast, non-essential activities, including retail, restaurants, entertainment, and self-care, contribute to enhancing quality of life but are not immediately critical for survival.

The period from August 1 through August 21, 2017, was designated as the pre-disaster phase. During this time, baseline levels for the four features were established using a seven-day moving average of daily visits to various POIs, and that level serves as a reference for comparing pre-disaster visitation patterns to POIs. This percentage change highlights the variance between daily visits to each POI category before and after the hurricane. In Equation 4, the daily value refers to the actual observed visits to POIs on a given day, while the base value acts as the reference point for the normal activity period before the disaster. The change percentage is then calculated to reflect variations in the daily visit numbers for each category, relative to the baseline, using a seven-day rolling average to smooth the data over seven-day intervals:

$$7 - day\ Rolling_{avg} = \sum_{d=1}^{7} \frac{Daily\ value - Base\ value}{Base\ Value} 100\% \qquad \text{(Eq. 4)}$$

Resilience curves for POIs were derived by plotting seven-day rolling averages. In this study, the term duration of recovery and is defined as the time at which the seven-day rolling average of visits to POIs attains 90% of the baseline values established prior to the disruption. This calculation enabled us to determine the average time needed for each area to achieve recovery.

## 4. Methodology

After computing the features related to the coupled human-infrastructure systems features, we implemented unsupervised machine learning to cluster CBGs based on their features. Figure 3 presents a detailed illustration of our analysis steps. The first step in our approach involved employing the PCA method to generate risk and resilience index and then applied Hierarchical cluster method to organize the two indexes into four distinct risk-resilience archetypes based on their feature values. The final step of analysis involved delineating features for each risk-resilience archetype that contributed to resilience of CBGs based on their risk levels. This analysis also included evaluating disparities in risk-resilience archetypes of CBGs with different median income levels.

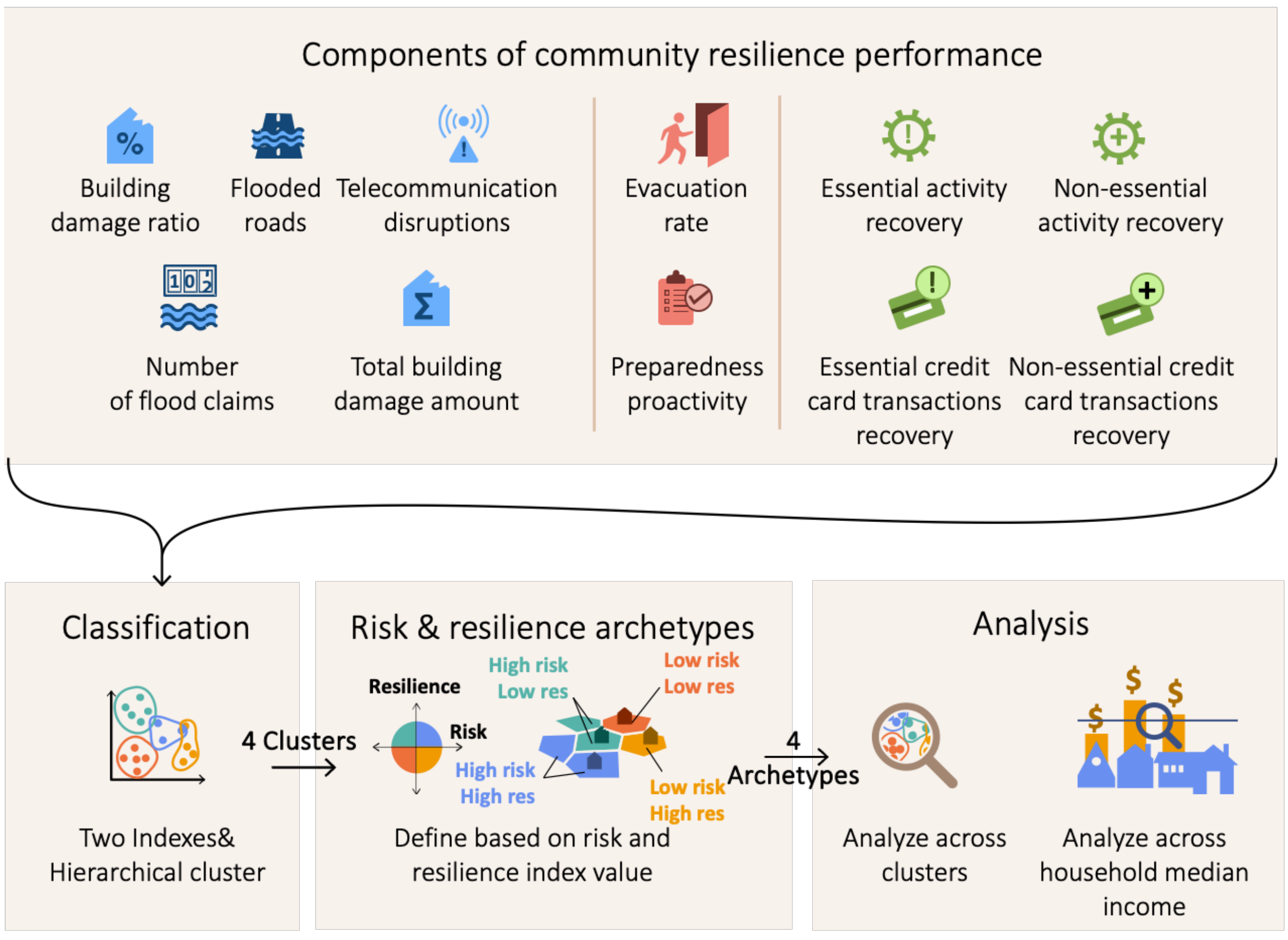

**Figure 3. The Overview of Analysis Steps.** Total of eleven features are classified into two indexes and then classified into four clusters using the Hierarchical algorithm. These clusters are then labeled as risk and resilience archetypes based on their index values. Accordingly, disparities in risk-resilience archetypes of CBGs with varying median income levels are examined.

### 4.2 Unsupervised classification based on Hierarchical Clustering

### 4.1 Data preparation

To quantify and understand patterns of risk and resilience across different regions, we employed a Principal Component Analysis (PCA) approach to reduce a set of carefully selected features into two composite indices—a Risk Index and a Resilience Index—based on a set of features related to risk and resilience ability. The Risk Index encompasses features including Evacuation Rate, Flooded Roads, Preparedness Proactivity, Number of Flood Claims, Damage Building Ratio, Total Building Damage Amount, and Telecommunication Disruptions. These features represent multiple dimensions of risk exposure and vulnerability within a region. The evacuation rate has consistently been a key element in risk and disaster management research (Sadiq, Tyler, & Noonan, 2019), and studying evacuation practices plays a crucial role in flood mitigation and reducing flood impacts. Flooded Roads highlight infrastructural vulnerabilities (Singh, Sinha, Vijhani, & Pahuja, 2018), as road flooding disrupts mobility, hampers emergency responses, and affects evacuation efforts, exacerbating the impact of disasters. Preparedness Proactivity measures the extent to which people take preparatory actions, such as visiting grocery stores or pharmacies before a hurricane. These behaviors provide insight into a community's readiness and awareness ahead of a disaster. Lower levels of proactive behavior may indicate greater vulnerability, which indirectly contributes to the overall risk profile in the Risk Index. The Number of Flood Claims serves as an indicator of the scale and frequency of flood damage, providing a quantitative measure of risk exposure and the financial burden on affected communities. The Damage Building Ratio and Total Building Damage Amount reflect the severity of structural damage caused by disasters, offering insight into the physical and economic vulnerabilities of a region. Finally, Telecommunication Disruptions capture the stability of communication networks during disasters, which are critical for coordination, information dissemination, and effective response. Together, these features provide a comprehensive understanding of the factors contributing to regional risk and vulnerability.

The Resilience Index, in contrast, focuses on features that reflect the speed of recovery, calculated as the reciprocal of recovery days for various activities. This index includes Essential Activity Recovery, Non-essential Activity Recovery, Essential Credit Card Transaction Recovery, and Non-essential Credit Card Transaction Recovery. To build the indices, we started by standardizing the data and transforming the resilience features, specifically by taking the reciprocal of recovery times for various activities to make shorter recovery times correspond to higher resilience. Then we calculated the covariance matrix of the standardized data, which captures the variance shared among pairs of features. This matrix reflects how features vary together, which is critical for identifying patterns that PCA can leverage. PCA then requires an eigenvalue decomposition of the covariance matrix. This decomposition yields eigenvalues $\lambda_i$ and eigenvectors $\omega_i$ , where each eigenvalue $\lambda_i$ quantifies the amount of variance explained by the corresponding principal component. The eigenvectors $\omega_i$ , also known as component loadings, define the direction of each principal component in the original feature space. The principal components themselves are linear combinations of the original standardized features, calculated as:

$$PC_k = X\omega_k \qquad \text{(Eq. 5)}$$

where $PC_k$ represents the k -th principal component, X is the matrix of standardized features, and $\omega_k$ is the eigenvector associated with the k -th largest eigenvalue. For each of the two indices, we selected only the first principal component, as it explains the largest proportion of variance in the data. The Risk Index and Resilience Index are then defined as:

$$\text{Risk Index} = X_{risk}\omega_{risk} \qquad \text{(Eq.6)}$$

$$\text{Resilience Index} = X_{resilience}\omega_{resilience} \qquad \text{(Eq.7)}$$

where $X_{risk}$ and $X_{resilience}$ represent the standardized feature matrices for the selected risk and resilience features, respectively, and $\omega_{risk}$ and $\omega_{resilience}$ are the first eigenvectors obtained from PCA on each feature set.

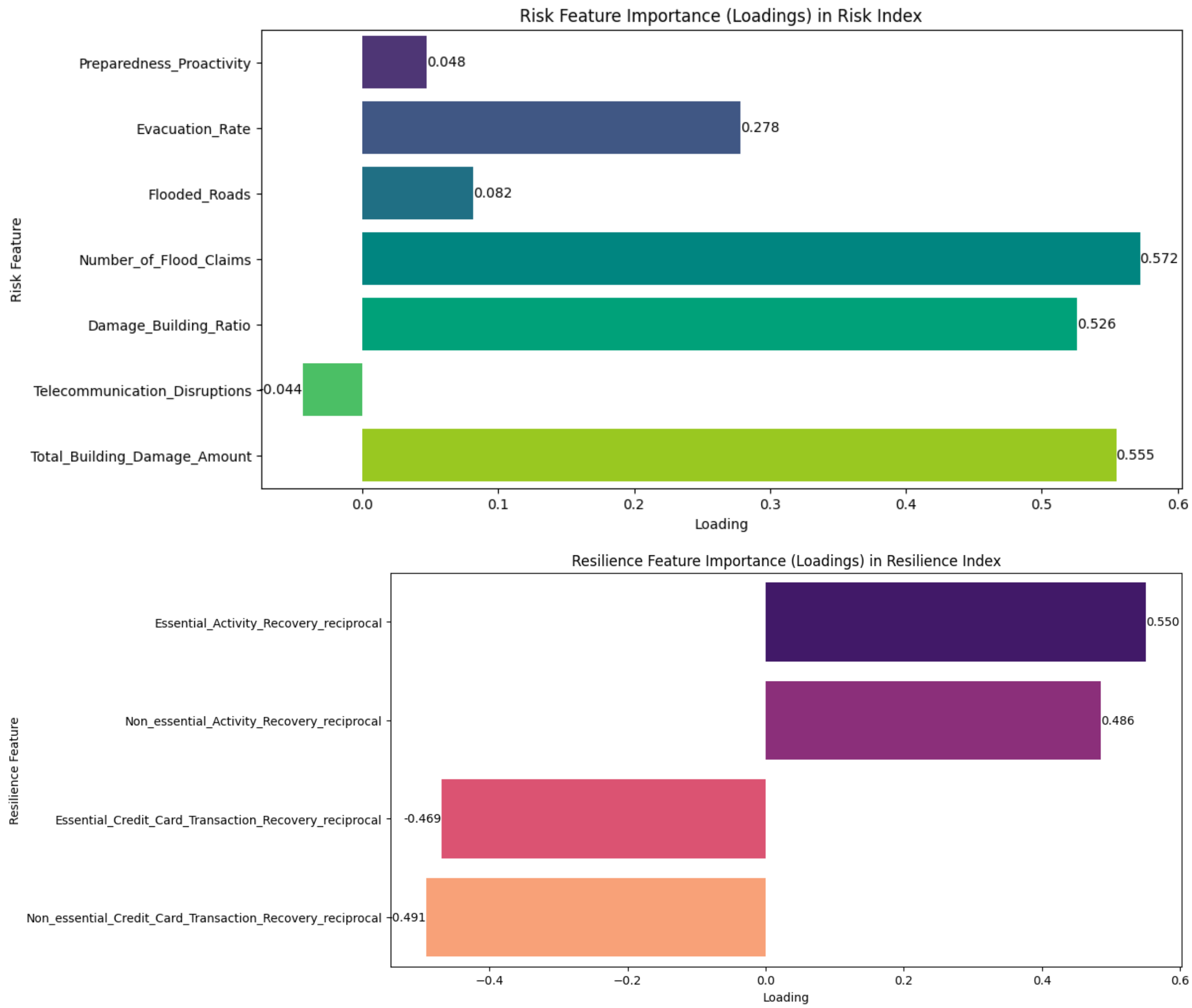

**Figure 4. Risk Feature and Resilience Features Importance to Index.**

Through the PCA loadings, we could see which specific features had the greatest influence on each index (Figure 4). For the Risk Index, the plot of feature loadings provides a clear view of the relative importance of each risk-related factor, revealing that "Number of Flood Claims" and "Total Building Damage Amount" are prominent, with loadings of 0.572 and 0.555, respectively. These features highlight the substantial role of physical damage and historical flood impact in shaping the overall risk profile of a region. Other features in the Risk Index, such as "Damage Building Ratio" (loading = 0.526) and "Evacuation Rate" (loading = 0.278), also show significant influence, although to a lesser extent. The moderate loading of "Evacuation Rate" suggests that behavioral responses to potential hazards contribute to risk but do not dominate the index. Additionally, features like "Flooded Roads" (loading = 0.082) and "Preparedness Proactivity" (loading = 0.048) have relatively lower weights, indicating their limited direct impact on the aggregated risk measure. "Telecommunication Disruptions" has a negative loading of -0.044, reflecting a slight inverse relationship, where increased telecommunication stability might correlate with higher perceived risk regions. In constructing the Resilience Index, we focused on features that indicate recovery efficiency across different sectors. The loading plot for resilience demonstrates that the reciprocal of "Essential Activity Recovery" duration has the highest influence with a loading of 0.550, underscoring its critical role in the resilience framework. This feature signifies how quickly essential activities return to baseline functionality, highlighting regions with robust recovery mechanisms. Similarly, the reciprocal of "Non-essential Activity Recovery" carries a substantial loading of 0.486, pointing to the importance of restoring non-essential services in evaluating resilience. The negative loadings observed for the reciprocal of "Essential Credit Card Transaction Recovery" (-0.469) and the reciprocal of "Non-essential Credit Card Transaction Recovery" (-0.491) reflect the impact of economic activity recovery patterns; regions where economic transactions resume slowly exhibit lower resilience scores.

To provide a comprehensive analysis, we also did feature relations to the Risk and Resilience indices using Ordinary Least Squares (OLS) regression results. According to the OLS results, the most influential predictor of the Risk Index is the Damage Building Ratio, with a coefficient of 4.933, signifying substantial weight on this indicator ($p < 0.0001$). Additionally, Total Building Damage Amount and Flooded Roads exhibit significant positive loadings (coefficients of 7.864 e-08 and 0.929, respectively, both significant at $p < 0.0001$). Telecommunication Disruptions exhibit a negative coefficient (-0.426), implying that areas with fewer disruptions tend to have lower risk scores. In the resilience index, key components include Essential Activity Recovery, Non-essential Activity Recovery, Essential Credit Card Transaction Recovery, and Non-essential Credit Card Transaction Recovery. According to the Ordinary Least Squares (OLS) regression results, Essential Activity Recovery and Non-essential Activity Recovery show positive coefficients (1.9067 and 1.7816, respectively, with $p < 0.0001$), indicating that quicker restoration of these services strongly enhances the Resilience Index. On the other hand, Essential Credit Card Transaction Recovery and Non-essential Credit Card Transaction Recovery have negative coefficients (-3.5164 and -4.4478, respectively, with $p < 0.0001$).

### 4.2 Hierarchical clustering

We then employed hierarchical clustering to identify patterns of risk and resilience index across regions. Hierarchical clustering is a technique that builds a multilevel hierarchy of clusters by either merging or splitting clusters iteratively. Specifically, we used the Agglomerative Clustering method, which is a "bottom-up" approach. Each data point starts in its own cluster, and pairs of clusters are merged as one moves up the hierarchy. Using the Ward linkage method, the agglomerative clustering merges clusters that minimize the within-cluster variance. This is expressed by calculating the variance criterion for each potential merge and choosing the pair of clusters that minimizes the increase in total within-cluster variance. The distance between clusters A and B in Ward's method can be formalized as:

$$D(A,B) = \frac{|A||B|}{|A|+|B|} \left|\left|\bar{x}_A - \bar{x}_B\right|\right|^2 \quad \text{(Eq. 8)}$$

where |A| and |B| are the sizes of clusters A and B , respectively, and $\bar{x}_A$ and $\bar{x}_B$ are the centroids of clusters A and B . This criterion tends to produce compact clusters with low variance, which aligns with the goal of identifying regions with distinct risk and resilience profiles. To decide the optimal number of clusters, we calculated the silhouette score for cluster numbers ranging from 2 to 9. Our silhouette analysis revealed that a four-cluster solution maximized the silhouette score at around 0.46, which, although moderate, is acceptable in the context of complex social and environmental data where overlapping clusters are common.

### 4.3 Risk and resilience archetypes

To visualize different archetypes, we calculate the mean value of risk and resilience index of all areas. The scatter plot in Figure 5 categorizes regions into four distinct clusters, each representing different combinations of risk and resilience characteristics within the framework of risk and resilience archetypes. The plot is organized with a horizontal dashed line marking the mean resilience value and a vertical dashed line representing the mean risk value, which helps to visually separate each cluster based on its mean risk and resilience index. The presence of confidence ellipses also highlights the internal heterogeneity of clusters. This segmentation reveals four quadrants that classify cluster 0 as low-risk and low-resilience (LL), cluster 3 as high-risk and low-resilience (HL), cluster 2 as low-risk and high-resilience (LH), and cluster 1 as high-risk and high-resilience (HH). We used a Chi-square test to examine the relationship between clusters and their distribution across four quadrants of risk and resilience, based on the overall mean values of the Risk and Resilience indices. The Chi-square test statistic is 1646.16, with a corresponding p-value of 0.0, indicating a statistically significant association between clusters and quadrants.

Cluster 0, represented in purple and positioned in the lower-left quadrant, combines low risk with low resilience (LL). With a mean Risk Index of -0.56 and a Resilience Index of -0.72, this cluster suggests areas that may not face severe risks but are also not good at recovery. Although they currently experience low risk, these areas remain vulnerable due to their limited recovery capacity. Cluster 1, shown in orange and located in the upper-right quadrant, represents regions that face high risk but have high resilience. With a high-Risk Index of 7.74 and a high Resilience Index of 0.31, Cluster 1 represents a high-risk, high-resilience (HH) group. This cluster identifies areas that, despite significant exposure to risk, have relatively strong recovery capabilities. The large confidence ellipse around Cluster 1's centroid suggests a broad variation within this group, indicating that while some areas might have moderate resilience, the overall cluster represents a segment that is predominantly vulnerable in the face of high risks. Cluster 2, depicted in blue and situated in the upper-left quadrant, stands out as a low-risk, high-resilience group (LH). With a mean Risk Index of -0.36 and a high Resilience Index of 1.18, this cluster represents regions that are well-prepared for recovery even though their risk exposure is relatively low. The tighter ellipse around Cluster 2's centroid indicates less variation within the cluster. Finally, Cluster 3, represented in green and positioned in the lower-right quadrant, combines high risk with low resilience (HL). With a mean Risk Index of 2.93 and a Resilience Index of -0.31, this cluster represents areas that are highly exposed to risk but exhibit limited recovery capacity.

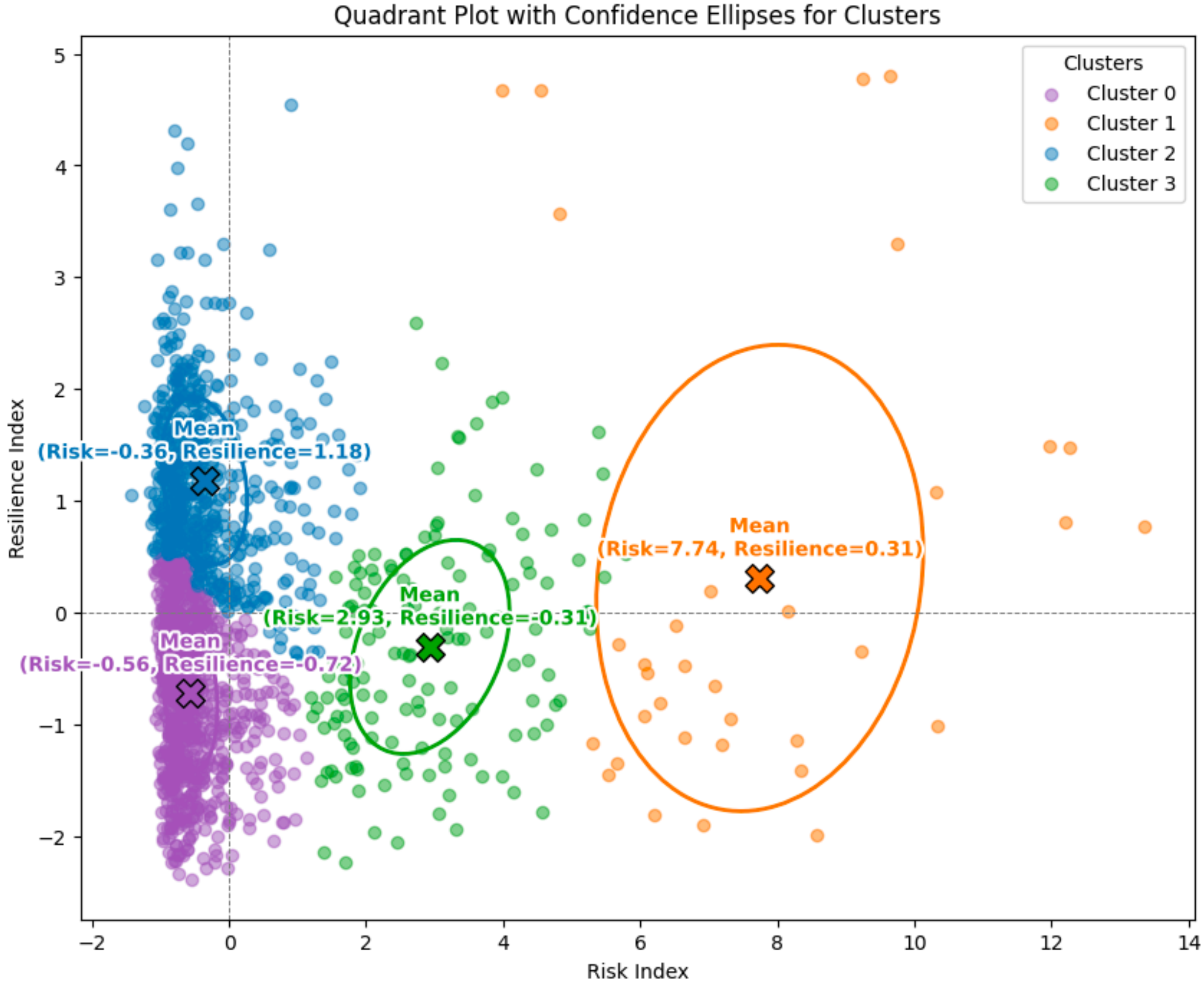

**Figure 5. Risk and resilience archetype for each cluster.** There are four resilience and risk archetypes: Cluster 0 is categorized under the low-risk/low-resilience (LL) quadrant; Cluster 1 is positioned in the high-risk/high-resilience (HH) quadrant; Cluster 2 falls into the low-risk/high-resilience (LH) quadrant; Cluster 3 is situated in the high-risk/low-resilience (HL) quadrant.

We have conducted the ANOVA results for both the Risk Index and Resilience Index further support significant differences across clusters: The F-statistic for the Risk Index is extremely high at 2630.70 with a p-value of 0.0, suggesting a clear and statistically significant difference in risk levels across clusters. The F-statistic for the Resilience Index is 665.47 with a p-value of 2.04 e-280, also indicating significant differences among the clusters in terms of resilience. This high level of significance highlights the variability in recovery capacities across clusters, reflecting how some regions can recover quickly while others cannot. We have also conducted the Tukey’s HSD test which provides detailed pairwise comparisons between clusters for both the Risk and Resilience Indices. For Risk Index, the mean difference between Cluster 0 and Cluster 1 is 8.31, with a p-value of 0.0, indicating that Cluster 1 has a significantly higher risk than Cluster 0. Cluster 0 and Cluster 3 differ by 3.49 ($p = 0.0$), with Cluster 3 exhibiting higher risk. Cluster 1 is significantly higher than both Cluster 2 (mean difference = -8.10) and Cluster 3 (mean difference = -4.81), highlighting that Cluster 1 has the highest risk profile. All pairwise comparisons in the Risk Index are statistically significant, indicating that each cluster exhibits a unique level of risk. For Resilience Index, Cluster 0 and Cluster 2 differ by 1.90, with Cluster 2 being more resilient. This difference is statistically significant ($p = 0.0$), underscoring that Cluster 2 represents high-resilience areas. The pairwise comparison between Cluster 1 and Cluster 3 reveals a mean difference of -0.62 ($p = 0.0003$), indicating that Cluster 3 has a significantly lower resilience than Cluster 1. All pairwise comparisons for the Resilience Index are statistically significant, confirming that each cluster has a distinct resilience level.

To assess whether key features related to risk and resilience vary significantly across the identified clusters, we conducted an Analysis of Variance ANOVA test, depending on the distribution of each feature. This approach allowed us to determine if the mean values of these features differ substantially between clusters, offering insight into whether distinct resilience and vulnerability profiles are present within each group. The analysis revealed that several features demonstrated statistically significant variation across clusters, with p-values well below the threshold of 0.05. For instance, both Essential Activity Recovery ($p \approx 2.69 \times 10^{-62}$) and Non-essential Activity Recovery ($p \approx 2.37 \times 10^{-44}$) exhibited highly significant differences, indicating diverse recovery capabilities among clusters. This suggests that clusters differ considerably in their ability to restore essential and non-essential services after a disaster, reflecting distinct resilience capacities. Similarly, credit card recovery features, including Essential Credit Card Transaction Recovery ($p \approx 2.95 \times 10^{-39}$) and Non-essential Credit Card Transaction Recovery ($p \approx 2.83 \times 10^{-42}$), displayed significant variation across clusters. Furthermore, the Evacuation Rate feature ($p \approx 9.15 \times 10^{-63}$) showed significant differences, suggesting that certain clusters are more proactive or capable in responding to disaster warnings. In addition, several risk exposure indicators exhibited significant variation across clusters. For example, Flooded Roads ($p \approx 6.6 \times 10^{-4}$), Number of Flood Claims ($p < 0.001$), Damage Building Ratio ($p < 0.001$), Telecommunication Disruptions ($p \approx 3.2 \times 10^{-2}$), and Total Building Damage Amount ($p < 0.001$) all differed significantly between clusters. In contrast, Preparedness Proactivity ($p \approx 0.157$) did not show statistically significant differences across clusters, as its p-value exceeded the threshold of 0.05. Most of the significant variation in recovery and risk exposure indicators across clusters suggests that these clusters capture meaningful differences in both vulnerability and resilience.

### 4.4 Spatial Difference of Four Archetypes

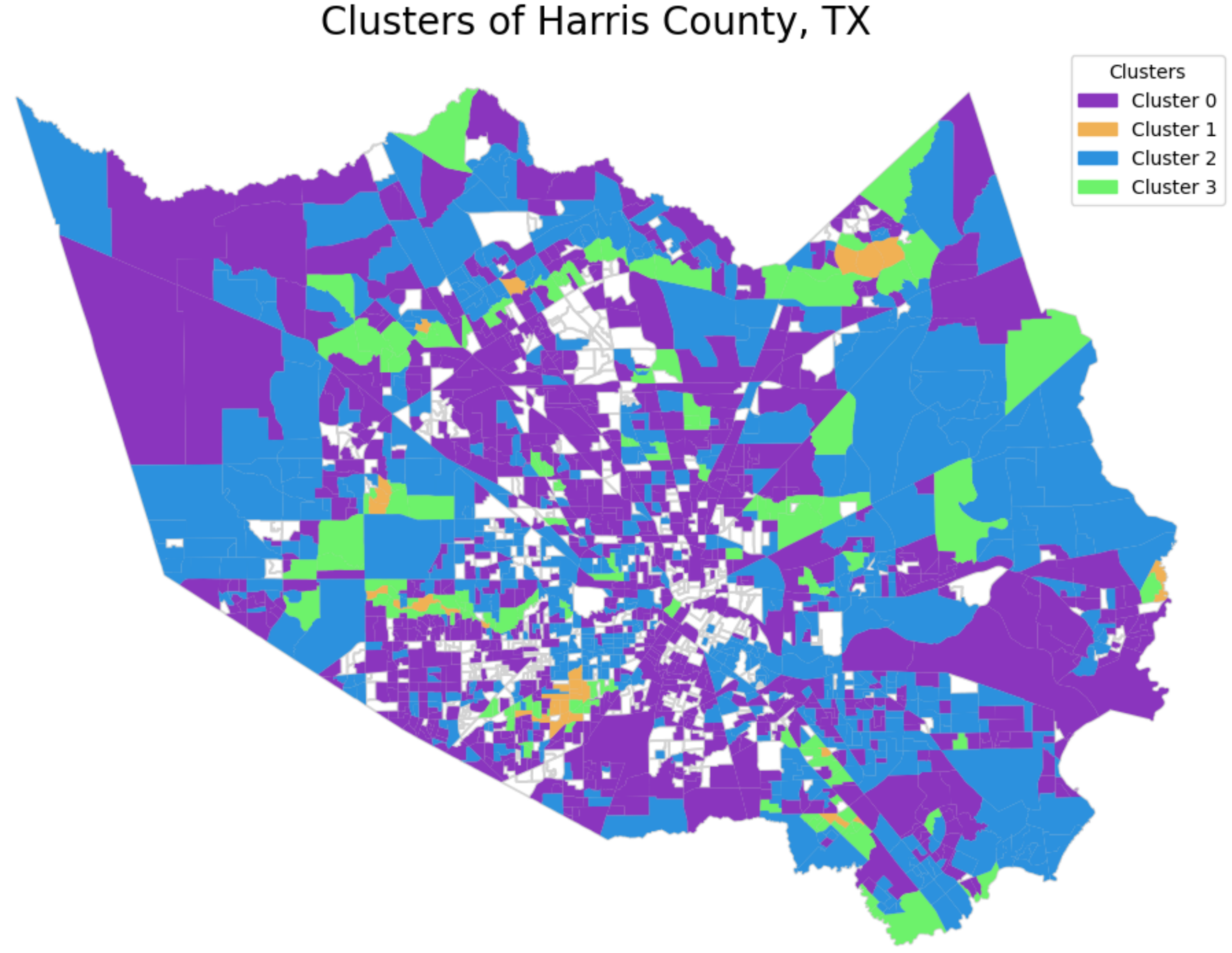

**Figure 6. Spatial Distribution of Clusters in Harris County, TX.** There are four clusters in Harris County based on the K-means clustering analysis: Cluster 0 (863 CBGs) is shaded purple, cluster 1 (34 CBGs), orange; cluster 2 (557 CBGs), blue; and cluster 3 (144 CBGs), green; and missing data, gray.

The map of Harris County, TX, displays the spatial distribution of the four identified clusters across the region, each represented by a unique color. Cluster 0 (purple) occupies a substantial portion of Harris County, particularly in the outer regions and sparsely populated areas. This cluster is characterized by low risk and low resilience (LL), which may indicate regions that are less urbanized and have limited infrastructure support for rapid recovery. The large coverage of Cluster 0 highlights its dominance in the county's less densely developed areas, potentially reflecting regions with limited disaster preparedness and recovery resources. Cluster 1 (Orange), identified with high risk and high resilience (HH), is relatively limited in its spatial distribution compared to the other clusters. Cluster 2 (Blue), defined by low risk and high resilience (LH), is scattered throughout Harris County but appears to be more prevalent in moderately urbanized zones. Cluster 3 (Green) represents low risk and low resilience (LL) and is found in various dispersed areas throughout the county. The distribution of Cluster 3 is relatively balanced, with a presence both near more developed areas and on the periphery.

## 5. Results and Discussion

### 5.1 Risk Features' relations to Resilience Index

We conducted an Ordinary Least Squares (OLS) regression to explore the relationship between risk features to the resilience index, aiming to understand how various factors of risk features influence resilience index across different regions (Table 2). Preparedness Proactivity showed a positive but statistically insignificant impact, indicating it may not be a decisive factor in enhancing resilience. In contrast, Evacuation Rate displayed a significant negative effect on resilience, suggesting that areas requiring frequent evacuations could face more complex or prolonged recovery. Flooded Roads, though positively associated with resilience, did not reach statistical significance, whereas the Number of Flood Claims had a positive and significant association with resilience, implying that areas with more frequent flooding may have developed stronger recovery systems or infrastructure support. Damage Building Ratio significantly reduced resilience, reflecting the substantial impact of structural vulnerabilities on delaying recovery. Telecommunication Disruptions, while positive, showed only marginal significance, indicating that communication issues might play a less critical role in resilience outcomes than other factors. Finally, Total Building Damage Amount had a small but significant negative effect, underscoring that extensive physical damage correlates with slower recovery. Although the model's $R^2$ of 0.038 suggests that risk features provide some insight into resilience, it also points to the influence of additional, possibly socio-economic or institutional, factors that might play more substantial roles in supporting recovery.

**Table 2. The relationship between Risk Features and Resilience Index**

| **Features** | **Coefficient** | **P-value** |
|---|---|---|
| Evacuation Rate | -0.1133 | 0.000 |
| Flooded Roads | 0.5167 | 0.122 |
| Preparedness Proactivity | 0.0200 | 0.322 |
| Number of Flood Claims | 0.0075 | 0.000 |
| Damage Building Ratio | -1.5593 | 0.000 |
| Total Building Damage Amount | -1.883e-08 | 0.030 |

| Telecommunication Disruptions | 0.5250 | 0.066 |
|---|---|---|

## 5.2 Features' Contribution to Index in Different Archetypes

To visualize the features' contribution to their own indexes, we applied the Clustered Feature Aggregation and examined the median values of various resilience and risk-related features across four distinct clusters (Figure 7). The resilience features (recovery-related features) are transformed using the reciprocal function (smaller values after transformation indicate better resilience). Median values of risk and resilience features are calculated within each cluster, and this step enables comparison across clusters, revealing patterns in how different clusters vary in their resilience and risk profiles. After aggregating the features within each cluster, Min-Max scaling is applied separately to the median values to bring all features into a comparable range (typically 0 to 1). This normalization facilitates a more interpretable comparison across clusters, regardless of the original feature scales.

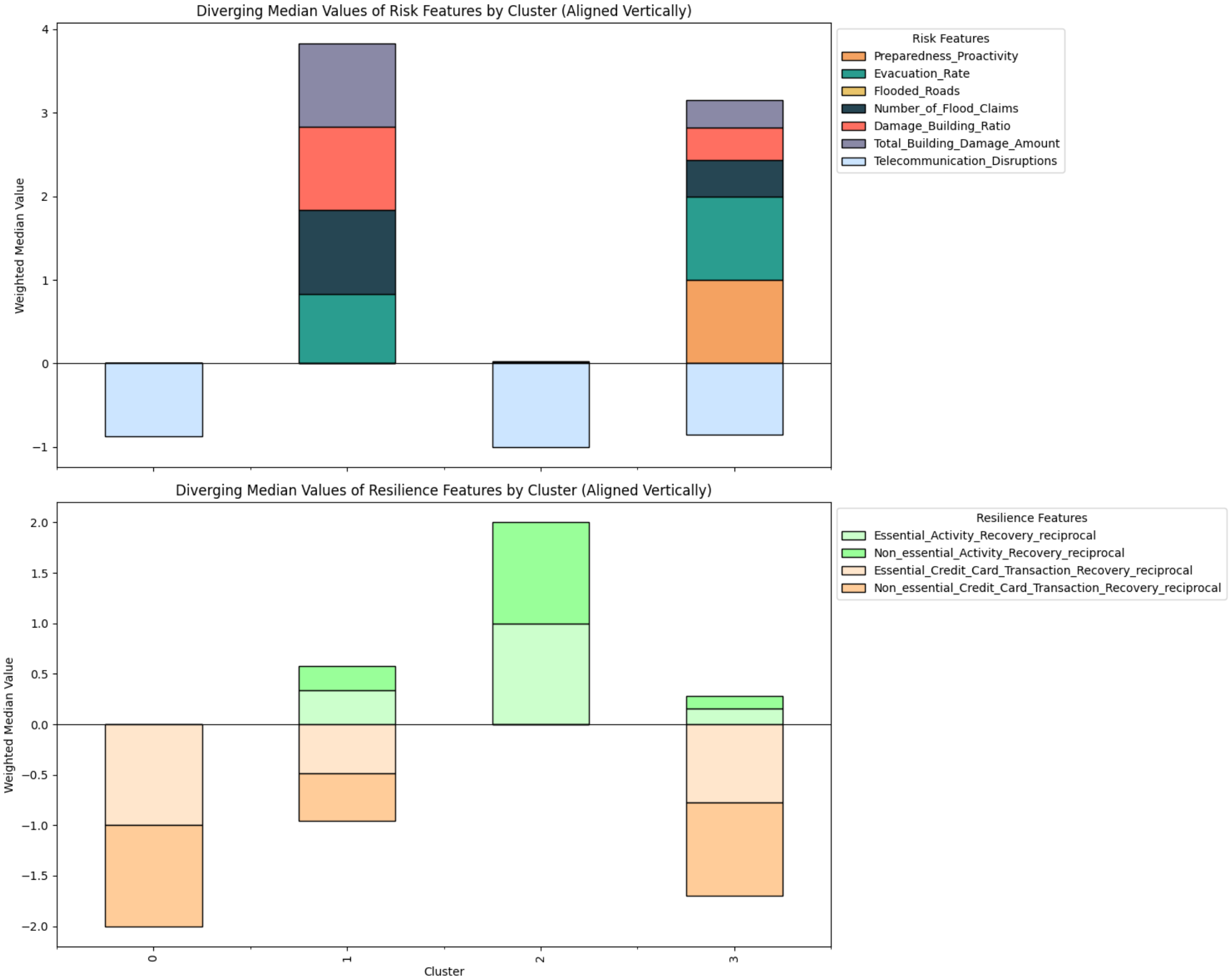

**Figure 7. Risk and Resilience Contributions.** Positive and negative contributions are visualized separately, with color-coding that enhances interpretability: positive resilience features are shaded in green, negative resilience features in orange, positive risk features are assigned random colors, and negative risk contributors, such as "Telecommunication Disruptions," appear in blue.

In Cluster 0, which is characterized by a low-risk and low-resilience (LL) profile, we observe minimal exposure to most risk factors, as evidenced by the relatively small, stacked values in the positive risk features. Telecommunication Disruptions is the primary risk feature for this cluster, but it is negative value in risk index, suggesting that telecommunication disruptions do not lead to high risk. In terms of resilience, Cluster 0 shows low values for essential and non-essential activity recover, but high values in non-essential card transactions. We know that card transaction is negative related to resilience, so Cluster 0 has lowest resilience compared to other clusters. Cluster 1, positioned in the high-risk, high-resilience (HH) quadrant, displays substantial values across several risk indicators, including high scores for "Total Building Damage Amount" and "Damage Building Ratio". These high values contribute significantly to the cluster's Risk Index, indicating that these areas are particularly vulnerable to physical and infrastructural risks. These two elements are also significantly negatively related to resilience index. Despite this vulnerability, Cluster 1 exhibits strong resilience, as seen in the high median values for essential and non-essential activity recovery. The resilience plot shows that activity recovery speed plays a crucial role in mitigating the impact of disruptions. Thus, although Cluster 1 faces considerable risk, its recovery capabilities allow it to effectively bounce back, particularly in critical services. Cluster 2 has a low-risk, high-resilience (LL) profile. This cluster shows low values across risk features, which like cluster 0. Resilience in Cluster 2 is particularly strong, as shown by highest values in essential and non-essential activity recovery, highlighting a robust capacity for restoring critical services after disruptions. Cluster 2 also shows minimal reliance on credit card recovery activities yet maintains a high Resilience Index due to strong activity recovery. Cluster 3 is characterized by high risk and low resilience (HL). The risk plot reveals significant exposure to risk factors, with high values for "Number of Flood Claims," "Building Damage Ratio," and "Total Building Damage Amount." These prominent risk factors contribute to its high-risk Index, indicating that this cluster is highly susceptible to disruptions. In terms of resilience, Cluster 3 shows lower scores for essential and non-essential activity recovery and negative scores for card transaction activities, reflecting limited recovery capabilities. Comparing Cluster 1 and Cluster 3, we observe that Cluster 1 has a higher "Total Damage Amount" than Cluster 3, leading to a higher risk index. This suggests that the damage amount serves as a significant indicator of an area's risk level. Cluster 2 has highest "Essential Activity Recovery" speed, and "Non-essential Activity Recovery" speed, also it does not have significant credit card activity, and this allow cluster 2 has highest resilience.

Overall, a key finding of this study is the significant role that the building system plays in determining the risk index. Features such as the building damage amount and building damage ratio are particularly influential, as higher values for these features correspond to an increased level of risk. This underscores the importance of infrastructure vulnerability in shaping the overall risk profile of a region. At the same time, the resilience index is strongly associated with the recovery of essential and non-essential activities. Higher performance in these activities—indicating a quicker return to normalcy after a disruptive event—captures greater resilience. This highlights the critical role of economic and social activity recovery in mitigating the long-term impacts of disasters and improving community adaptability. By linking the building system to risk and activity recovery to resilience, our study emphasizes the interconnected nature of physical infrastructure and social dynamics in disaster management.

### 5.3 Disparities in risk and resilience archetypes across income groups

In order to analyze the relationship between median income and the four archetypes, we used both descriptive statistics and statistical testing to explore how income varies across the clusters, as well as to determine if there is a statistically significant difference in median income among them. The box plot of median income by cluster visually highlights the distribution of income within each cluster, allowing for a detailed inspection of central tendency and spread.

From the box plot, we observe that Cluster 0 (colored in purple) has a median income of approximately $54,618 and contains the largest number of areas with a count of 845. This cluster, with a relatively lower median income, represents areas with significant dispersion and several low-income outliers, indicating economic diversity. Cluster 1 (in orange) has the highest median income of about $113,380, albeit with a much smaller sample size of 34 areas. Cluster 2 (blue) and Cluster 3 (green) have similar median incomes, around $61,924 and $62,962 respectively, with 552 and 142 areas in each cluster.

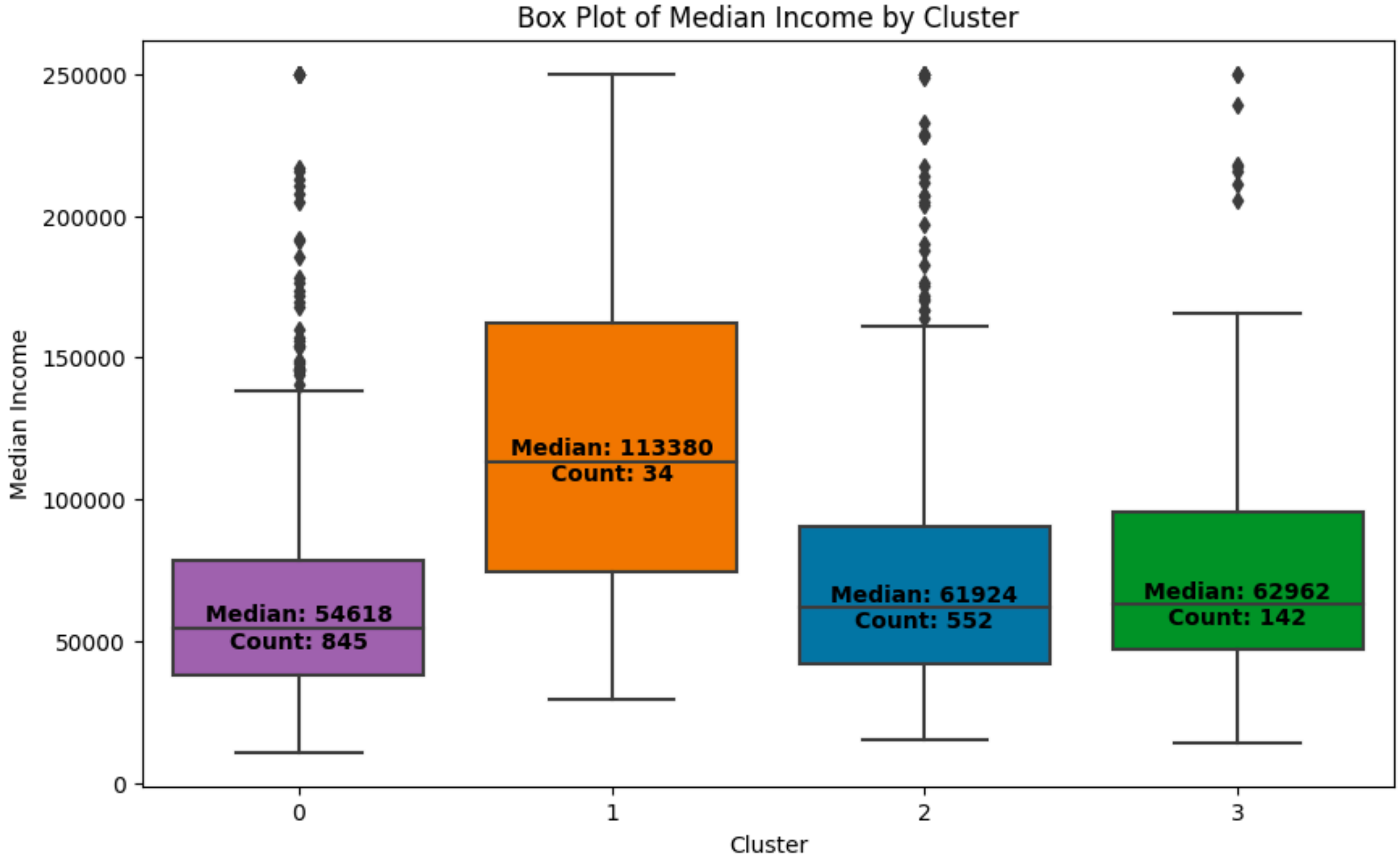

**Figure 8. Box Plot of Median income by Cluster**

To statistically validate these observations, we conducted a one-way ANOVA test, which allows us to assess whether there are significant differences in median income across the clusters. The ANOVA results indicate a significant effect of cluster membership on median income ($p < 0.001$), confirming that the median income levels differ across the clusters in a statistically meaningful way.

Additionally, we investigated the relationship between median income and resilience, accounting for the influence of risk. To do this, we used a multiple regression model with the Resilience Index as the dependent variable and Median Income, Risk Index, and their interaction as predictors. The regression equation can be generally represented as follows:

$$Y = \beta_0 + \beta_1 X_1 + \beta_2 X_2 + \beta_3 (X_1 \times X_2) + \varepsilon \quad \text{( Eq.10)}$$

where $Y$ is the Resilience Index, $X_1$ represents Median Income, $X_2$ is the Risk Index, $X_1 \times X_2$ denotes the interaction between Median Income and Risk Index, and $\varepsilon$ is the error term. The coefficients $\beta_1$ ,$\beta_2$ ,and $\beta_3$ provide insights into the influence of each variable and their interaction on resilience.

The results from the regression analysis show that the Median Income coefficient is positive and statistically significant ($4.756 \times 10^{-6}$, $p < 0.001$), suggesting that higher median income is associated with higher resilience, likely due to the economic resources available for recovery. The Risk Index has a negative coefficient (-0.0928, $p = 0.010$), indicating that higher risk levels are associated with lower resilience, independent of income. Importantly, the interaction term between Risk Index and Median Income ($7.546 \times 10^{-7}$, p = 0.012) is also significant, which implies that income moderates the relationship between risk and resilience. Higher-income areas appear to experience a less severe negative impact of risk on resilience, suggesting that economic resources may buffer some of the detrimental effects of high-risk exposure.

In summary, these findings indicate that while median income positively affects resilience, risk remains a limiting factor. The presence of a significant interaction term highlights that the relationship between risk and resilience is moderated by income, where wealthier areas are better equipped to handle risks, thus mitigating their impact on resilience. This analysis underscores the critical role of socio-economic factors in influencing resilience outcomes and suggests that income-based interventions could enhance resilience in high-risk, lower-income regions.

### 5.4 Interaction of Risk Index and Median Income on Resilience

We used Ordinary Least Squares (OLS) regression to examine the influence of the interaction between risk index and median income on resilience index across four distinct archetypes. For each cluster, we introduced an interaction term, Risk_Median_Interaction, calculated as the product of the Risk Index and Median income, to evaluate the compounded effect of these variables on Resilience Index.

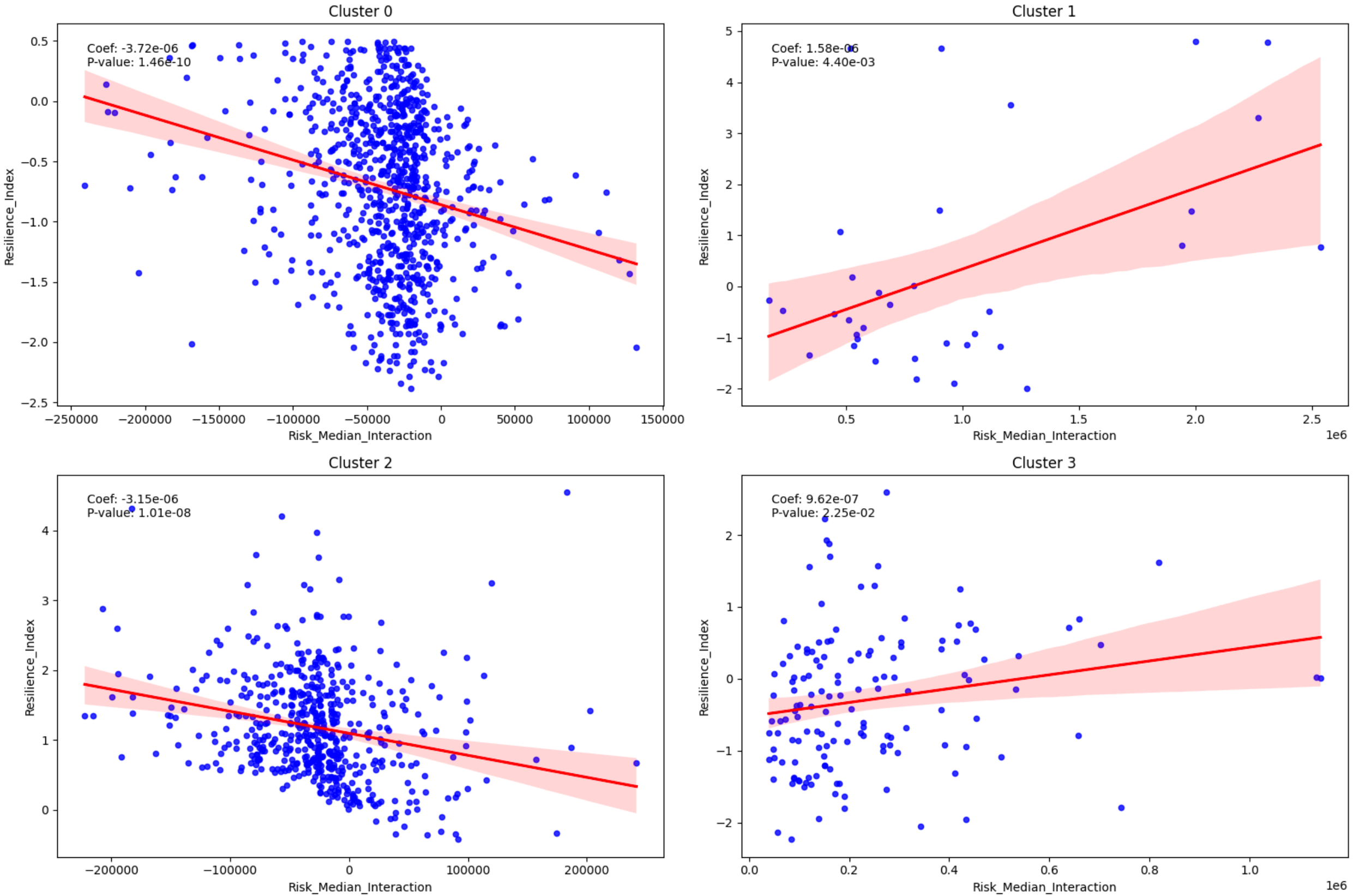

**Figure 9. Relationship between Risk-Median Income and Resilience Index**

In Figure 9, we can see that Cluster 0 shows a highly significant, negative coefficient for the interaction term (-3.717e-06, $p < 0.001$). This result suggests that in this cluster, characterized by the lowest median income level, areas with both high risk and higher income tend to experience lower resilience. It is possible that the relatively low baseline income in Cluster 0 limits the effectiveness of income as a resilience buffer in high-risk scenarios. In Cluster 1, which has the highest median income among all clusters, the interaction term's coefficient is positive (1.583e-06) and statistically significant ($p = 0.004$). This finding indicates that in wealthier regions, higher income in high-risk areas is associated with increased resilience. The positive interaction suggests that income contributes meaningfully to resilience in this cluster, likely due to enhanced capacity for investments in infrastructure, insurance, and community resources that can buffer against risk. These high-income areas are thus better equipped to withstand and recover from disruptions, underscoring the role of financial resources in risk management and community resilience. For Cluster 2, the interaction effect is again significantly negative (-3.153e-06, $p < 0.001$), mirroring the trend observed in Cluster 0. This cluster has a second lowest income level, and the results suggest that while income may provide some advantages, it does not fully counteract the negative influence of high risk on resilience. The findings here imply that in low-income areas, resources may not be effectively allocated to resilience-enhancing activities, or that socio-economic factors limit the resilience benefits of higher income. Cluster 3, which has the second-highest income level, shows a positive interaction effect (9.623e-07) with a p-value of 0.022. Although the effect is smaller than in Cluster 1, this result still suggests that higher income has a modestly positive influence on resilience in high-risk contexts within this cluster. The moderate positive interaction may reflect that areas in Cluster 3 are investing in some resilience-supporting resources, though not as extensively as those in Cluster 1.

The findings from each cluster emphasize that the impact of the interaction between risk and Median income on resilience is context dependent. Clusters with higher baseline income levels, such as Clusters 1 and 3, demonstrate a positive association between income and resilience in high-risk areas, likely due to more substantial investments in resilience resources. Conversely, lower-income clusters like Clusters 0 and 2 exhibit a negative interaction effect, indicating that higher income alone does not suffice to offset the challenges posed by high risk. Overall, this analysis reveals that income has a complex and context-specific influence on resilience in high-risk environments. Higher-income areas have a more protective response to risk, while lower-income areas may struggle to achieve resilience despite high income levels. This underscores the need for tailored strategies that consider both income and risk levels to enhance resilience effectively across diverse communities.

## 6. Concluding Remarks

This study proposed a machine learning-based approach for ex-post assessment of community risk and resilience based on coupled human-infrastructure systems performance. Despite significant advancements in the interdisciplinary field of community resilience, most existing studies focus primarily on ex-ante assessments; Relatively less attention has been paid to ex-post assessments. Also, many ex-post assessments are based primarily on disaster reconnaissance approaches focusing on measuring impacts and damage. Furthermore, ex-post assessments rely primarily on surveys for capturing impacts and damage, resulting in delays in data collection and assessment. Finally, the current approaches to ex-post assessment of community resilience focus mainly on single systems (e.g., transportation, housing, or businesses) and do not consider various aspects of coupled human-infrastructure performance. These limitations are addressed in this study based on evaluating various features related to coupled-human infrastructure performance computed from various data sources and by adopting a machine learning-based approach to classify spatial areas of a community (census block groups) based on the intertwined features to unveil various risk and resilience performance archetypes.

The outcomes of this study have multiple important scientific and practical contributions. First, the study provides a novel approach to ex-post assessment of community resilience using a data-driven and machine intelligence-based approach. The method presented in this study could provide interdisciplinary researchers in the fields of disasters and community resilience with a new tool for ex-post evaluation of community resilience based on a diverse range of features related to coupled human-infrastructure systems performance. This data-driven approach could provide a more unified way for ex-post evaluation of community risk and resilience to enable comparison of coupled human-infrastructure systems performance across different events and different regions. Second, the approach presented in this study enables capturing various features related to coupled human-infrastructure performance in evaluation of community resilience. The features capture disruptions and restoration of infrastructure services, protective actions of populations, property damage, as well as life activity recovery of populations. The consideration of various features would provide a more comprehensive evaluation of community resilience compared with the existing approaches that focus primarily on performance of human or infrastructure systems separately.

Third, the features related to the coupled human-infrastructure systems performance are captured based on observational data from various sources enabling a more data-driven approach to post-disaster community resilience evaluations. Data-driven evaluation community resilience based on various coupled human-infrastructure systems performance features is essential for guiding response and recovery efforts to fairly effect resource allocation and prioritization. Fourth, the study revealed the spatial risk and resilience profile of communities based on four archetypes according to their coupled human-infrastructure systems performance. The specification of the risk-resilience archetype of communities provides essential insights for future risk reduction and resilience improvement strategies for planners and public officials. For example, generally, essential and non-essential activity recovery are more reasonable to represent the positive resilience conditions for ex-post assessment. Higher activity recovery represents higher resilience. In risk features, preparedness proactivity, evacuation rate, flooded roads, number of flood claims, total building damage amount, and building damage ratio all positive contribute to risk index. This suggests that these clusters frequently face evacuation scenarios and endure significant structural damage during events, indicating an urgent need for infrastructure resilience improvements. Investments in structural fortification and enhanced evacuation protocols would reduce the frequency and severity of these disruptions, supporting faster recovery times and greater resilience in these high-risk areas. In contrast, strong resilience, particularly in Essential and Non-essential Activity Recovery, suggests that it has robust systems in place to resume key activities after disruptions. In addition, features like Total Building Damage Amount play a significant role in determining an area's risk levels. For example, a higher Total Building Damage Amount during disaster events consistently correlates with increased risk exposure, as observed across different datasets. This finding allows us to identify specific features that lead to heightened risk, offering insights into patterns of vulnerability across diverse events. Such insights can be instrumental for governments and policymakers in prioritizing infrastructure strengthening initiatives. For instance, investing in building retrofitting, enhanced construction standards, and damage mitigation strategies could significantly reduce risk in vulnerable areas. By understanding the role of building damage in driving risk, communities can implement targeted measures to minimize physical losses and bolster disaster preparedness. Similarly, Non-essential Activity Recovery emerges as a critical driver of resilience. This feature represents the community's ability to restore non-essential economic and social activities, such as retail, entertainment, and leisure, which are essential for long-term recovery and economic stability. For example, areas demonstrating faster recovery in non-essential activities tend to exhibit greater overall resilience in the aftermath of disasters. These findings provide actionable insights for governments and local organizations to focus on enabling quicker restoration of non-essential services. Strategies such as ensuring faster utility restoration, supporting small businesses, and creating flexible recovery plans for non-essential sectors can significantly enhance community resilience. Fifth, the findings of the analysis in the context of Harris County in the 2017 Hurricane Harvey revealed disparities in the risk-resilience profile of different median income communities. The results reveal that higher baseline income levels can enhance resilience in high-risk areas, likely due to the ability to invest more substantially in resilience resources. However, this positive association is not universal; lower-income areas, even those with relatively high-income levels within their context, may still struggle to achieve resilience when faced with significant risk factors. This suggests that income alone does not uniformly confer resilience, particularly in communities exposed to elevated risk levels. Therefore, to effectively enhance resilience across diverse communities, tailored strategies are essential strategies that take into account not only income levels but also the unique risk profiles of each community.

The study outcomes also set the stage for future studies to advance community resilience in two main directions. First, future studies can include additional features related to the coupled human-infrastructure systems performance when data is available. Second, the approach presented in this study can be adopted for ex-post community resilience assessment across different events (in the same region) or across different regions to consistently compare the performance of coupled human-infrastructure systems across events and regions and their contributions to community resilience. Through these cross-regional and cross-events ex-post assessments, important features that positively contribute to resilience can be identified.

**Data Availability**

The data that support the findings of this study are available from Ookla, Safegraph, INRIX, Microsoft Building Footprint, and Spectus, Inc, but restrictions apply to the availability of these data, which were used under license for the current study. The data can be accessed upon request. Other data we use in this study are all publicly available.

**Code Availability**

The code that supports the findings of this study is available from the corresponding author upon request.

**Acknowledgement**

The authors would like to acknowledge funding support from the NSF CAREER Faculty Early Career Development Program. The authors also would also like to acknowledge the data support from Ookla, Safegraph, INRIX, Microsoft Building Footprint, and Spectus, Inc. Any opinions, findings, conclusions, or recommendations expressed in this material are those of the authors and do not necessarily reflect the views of Ookla, Safegraph, INRIX, and Microsoft building footprint, and Spectus, Inc.